\documentclass[eqsecnum,aps,prd,nofootinbib]{revtex4}
\usepackage[dvips]{graphicx}
\usepackage{amsmath,amssymb,graphicx,mathrsfs}
\usepackage{epsfig}

\newcommand{\be}{\begin{equation}}
\newcommand{\ee}{\end{equation}}
\newcommand{\bea}{\begin{eqnarray}}
\newcommand{\eea}{\end{eqnarray}}
\newcommand{\bn}{\bar{\nabla}}
\newcommand{\p}{\partial}
\newcommand{\bg}{{\bar g}}
\newcommand{\G}{{\mathcal
 G}}
\newcommand{\bxi}{{\bar \xi}}
\newcommand{\bnabla}{{\bar \nabla}}

\newcommand{\sL}{\mathscr L}

\def\({\left(} \def\){\right)}

\def\revise#1       {\raisebox{-0em}{\rule{3pt}{1em}}
                   \marginpar{\raisebox{.5em}{\vrule width3pt\
                     \vrule width0pt height 0pt depth0.5em
                  \hbox to 0cm{\hspace{0cm}{%
                     \parbox[t]{4em}{\raggedright\footnotesize{#1}}}\hss}}}}

\usepackage{hyperref}

\begin{document}
\title{Wald-like formula for energy}

\author{Aaron J. Amsel}
 \email{aamsel@asu.edu}
\affiliation{Department of Physics and Beyond Center for Fundamental Concepts in Science, \\ Arizona State University, \\Tempe, AZ 85287}

\author{Dan Gorbonos}
\email{dan.gorbonos@mail.huji.ac.il} \affiliation{no affiliation}

\begin{abstract}
We present a simple ``Wald-like'' formula for gravitational energy
about a constant-curvature background spacetime.  The formula is
derived following the Abbott-Deser-Tekin approach for the definition
of conserved asymptotic charges in higher-derivative gravity.
\end{abstract}

\maketitle

\tableofcontents

\section{Introduction}
In this paper, we present a simple formula for the computation of
global charges (in particular energy) in asymptotically constant-curvature spacetimes for general gravitational theories. This is of
particular interest for solutions of higher-derivative theories that
approach  a constant curvature solution in the asymptotic region.

Our formula is similar to Wald's formula for
entropy~\cite{wald1,wald2,myers} in the sense that both formulas
involve a derivative of the Lagrangian of the theory with respect to
the Riemann tensor. The main difference is that the entropy is
computed as an integral over the horizon, while the energy is
computed in the asymptotic region (where we regard the solution as a
perturbation of the background). Another difference is that Wald's
formula involves only first derivatives with respect to curvature,
whereas the formula for energy involves second derivatives.
Nevertheless, the two formulas should be related by the first law of
black hole thermodynamics and its integrated forms, i.e.,~Komar
integrals and the Smarr formula [in some extended form to higher-derivative gravity and (A)dS spacetimes; see,
e.g., Refs.~\cite{Kastor,Kastor2}]. Proposals for similar Wald-like
formulas for the shear viscosity were given in Refs.~\cite{shear1,shear2}.

This approach gives us a new viewpoint on black hole thermodynamics.
Wald's entropy formula was reinterpreted as the Bekenstein-Hawking
entropy with an effective gravitational coupling~\cite{entropy} (see
also Ref.~\cite{entropy2}). This effective coupling comes from the
coefficient of the kinetic term of a specific type of metric
perturbation in the theory. Equivalently, this specific type of
perturbation corresponds to the propagator for the exchange of a
graviton between two covariantly conserved sources at the horizon,
and the effective coupling comes from this propagator. Thus, Wald's
formula not only has an advantage in the computational aspect, but
also gives us a microscopic interpretation of the black hole
entropy.

In a similar way, a Wald-like formula for energy can be naturally
interpreted as giving the effective gravitational coupling in the
asymptotic region, namely, on the background. This avenue was
explored in Ref.~\cite{Tekin} for cubic corrections and
in Ref.~\cite{Sisman:2012rc} for Lanczos-Lovelock gravity, where the effective
gravitational coupling was identified as coming from
higher-derivative corrections to the tree-level scattering amplitude
between two background covariantly conserved sources via the
exchange of a graviton. This is the same effective coupling that
appears in the Wald-like energy formula derived below. This
viewpoint, for both the entropy and energy, gives us two sides of
the microscopic interpretation (in terms of corrections to a
graviton exchange amplitude) of higher-derivative corrections to
black hole thermodynamics.

The derivation of the formula is based on the
Abbot-Deser-Tekin (ADT) method for computing energy
 \cite{Deser:2002rt,Deser:2002jk}. We basically reduce
their general method to a single formula which only requires
substitution of the Lagrangian and the background solution. The
story of this method starts with the result of Arnowitt, Deser, and Misner (ADM) \cite{ADM} for
energy in Einstein-Hilbert gravity with asymptotically flat boundary
conditions. This was later generalized to spacetimes with a
cosmological constant in Ref. \cite{Abbott:1981ff}. These so-called ``Abbott-Deser (AD)
charges'' were written in a manifestly covariant way and once again
could be expressed as pure surface integrals.  The method used to
construct the AD charges was then further generalized to arbitrary
higher-curvature theories in Refs. \cite{Deser:2002rt,Deser:2002jk}.

As we will see, the ADT method involves relatively little formalism
and is computationally straightforward. In addition, this method has
the advantage of not involving any explicit regularization or
subtraction of infinities, as required in counterterm methods (see,
e.g., Refs. \cite{Henningson:1998gx,Balasubramanian:1999re}). Unlike
Euclidean path integral techniques (e.g., \cite{Hawking:1980gf}), the
ADT framework naturally gives the gravitational mass as an integral
at asymptotic infinity, without any need to identify a horizon in
the interior.  For perturbations that vanish sufficiently fast at
asymptotic infinity, the ADT
charges are exactly the same as the charges derived using the
covariant phase space methods of
Refs. \cite{Barnich:2001jy,Barnich:2007bf,Compere:2007az}, which in turn
differ from the charges of Wald et al.
\cite{Lee:1990nz,wald2,Wald:1999wa} by a surface term proportional
to the Killing equations.

This paper is organized as follows.  In Sec.~\ref{energyformula},
we present the Wald-like formula for energy and
explain how to use it.  We also give  the energy of black hole solutions for two examples:
Gauss-Bonnet gravity, and a theory with six-derivative
corrections that was previously studied in Ref.~\cite{previous}. We show that the calculation of energy becomes much
shorter and simpler when using the formula.  The rest of the paper is devoted to deriving this formula from
the ADT method.  After a short presentation of the ADT method in
Sec.~\ref{ADT}, we discuss the general structure of the ``effective''
stress-energy tensor in higher-derivative gravity in Sec.~\ref{flat}.  We then obtain an explicit formula for the
stress-energy tensor for a flat background in Sec.~\ref{flat2}.  This expression is generalized to a curved background in Sec.~\ref{energy-momentum}, using the effective quadratic curvature method \cite{Tekin,Hindawi:1995cu,Gullu:2010em,Gullu:2010vw,Sisman:2012rc}.   Following Ref.~\cite{Deser:2002rt}, in
Sec.~\ref{the derivation} we complete the computation by deriving the energy from the general
expression of the stress-energy tensor. We conclude with
a brief discussion of our results and future directions in
Sec.~\ref{conclusions}.

\section{The Formula for Energy}
\label{energyformula}

Here we present the formula for energy and explain how to use it.
Let us consider a general $d$-dimensional theory of gravity whose
action depends on the metric $g_{\mu\nu}$ and the curvature (through
the Riemann tensor)

\be I=\int d^{d}x \sqrt{-g}\,
\mathscr{L}(R_{\mu\nu\rho\sigma},g_{\mu\nu})\,.\label{general
lagrangian}\ee
We will construct the energy for such theories following the approach of Refs. \cite{
Abbott:1981ff,Deser:2002rt,Deser:2002jk}.  We assume that the action is invariant under
diffeomorphisms. In order to define a gauge-invariant conserved
charge we need the presence of an asymptotic Killing symmetry.  The charge is then defined relative to a background solution, denoted as $\bg_{\mu\nu}$, which admits a Killing vector $\bxi_{\mu}$.  We assume that the background is a homogeneous solution, namely, described by an ``effective'' cosmological constant $\Lambda$,
which can be negative, positive, or zero (which is the asymptotically
flat case). In addition, the solutions are required to fall off sufficiently fast at infinity relative to the background\footnote{Here we mean that the perturbation
about a solution falls off fast enough at infinity that the theory
is asymptotically linear, i.e. that the linearized equations of
motion are obeyed near infinity.  In this case, the charges of the
linearized theory can be used to obtain the charges of the
nonlinear theory.  For example, this condition holds for the case of
standard asymptotically AdS boundary conditions
\cite{Henneaux:1985tv}.}.

For a large class of solutions [which includes the asymptotically
Schwarzschild-(A)dS (SdS) solutions defined below], we can write the energy of a generic
higher-derivative theory in the same form as applies to
Einstein-Hilbert gravity (with a cosmological constant), but with an
overall multiplicative factor that depends on the higher-curvature
terms. This will later serve as a basis for the interpretation of
higher-derivative corrections as effective modifications of the
gravitational coupling constant (Newton's constant). The Lagrangian
of Einstein-Hilbert gravity with a cosmological constant is \be
\label{Einstein action}
\mathscr{L}_{E}=\frac{1}{2\kappa}\(R-2\,\Lambda_0\), \ee where
$\kappa$ is the $d$-dimensional gravitational coupling constant. The
ADT energy for solutions of this theory is denoted by $E_0$ and is
given explicitly in Eq. (\ref{surface Einstein}). Then, the energy
for the general Lagrangian (\ref{general lagrangian}) is \be
\label{ENERGY FORMULA}
\boxed{E=\left[P_{\rho\sigma}^{\mu\nu}\(\frac{\partial \mathscr{
L}}{\partial
R^{\mu\nu}_{\rho\sigma}}\)_{\bar{g}}-\frac{4\,\Lambda(d-3)}{(d-1)(d-2)}P^{\phantom{(2)}\gamma\delta,\mu\nu}_
   {(1)\alpha\beta\phantom{,}\rho\sigma}\(\frac{\partial^{2}\mathscr{
L}}{\partial R^{\gamma\delta}_{\alpha\beta} \partial
R^{\mu\nu}_{\rho\sigma}}\)_{\bar{g}}\right] 2 \kappa E_0\,,}
 \ee
where
 \bea
P_{\rho\sigma}^{\mu\nu}&=&\frac{2\,\delta_{[\rho}^\mu \delta_{\sigma]}^\nu}{d(d-1)},\label{projection0}\\
\label{projection1} P^{\phantom{(1)}\gamma\delta,\mu\nu}_
   {(1)\alpha\beta\phantom{,}\rho\sigma}&=&\frac{4}{d(d^2-1)(d-2)(d^2-2\,d+2)}
\left[(d-1)^2\delta^{\mu}_{[\alpha } \delta^{\nu}_{\beta]}
\delta_{\rho}^{[\gamma } \delta_{\sigma}^{\delta]}
-\delta^{\gamma}_{[\alpha } \delta^{\delta}_{\beta]}
\delta_{\rho}^{[\mu } \delta_{\sigma}^{\nu]}
-(d-2)\delta^{\delta}_{[\beta } \delta^{\mu}_{\alpha]}
\delta_{[\sigma}^{\nu} \delta_{\rho]}^{\gamma} \right]\,.
 \eea
[The complete energy formula for more general boundary conditions is given in Eq. \eqref{g charge}.] Here $\Lambda$ is the effective cosmological constant associated with the background solution $\bg_{\mu \nu}$, which in general is distinct from the ``bare'' cosmological constant $\Lambda_0$ that may appear in the action.

The derivative of the Lagrangian with respect to
$R^{\mu\nu}_{\rho\sigma}$ is performed formally, as if
$R^{\mu\nu}_{\rho\sigma}$ and $g_{\mu\nu}$ are independent, and we
impose the same tensor symmetries as $R^{\mu\nu}_{\rho\sigma}$. For
example, in the case of Einstein gravity (\ref{Einstein action}) we
get \be \frac{\partial \mathscr{L}_{E}}{\partial
R^{\mu\nu}_{\rho\sigma}}=\frac{1}{2
\kappa}\delta^{\rho}_{[\mu}\delta^{\sigma}_{\nu]}. \ee The
$(\ldots)_\bg$ notation indicates that the expression in parentheses
is to be evaluated on the background spacetime $\bg_{\mu \nu}$.

The expressions in Eqs. (\ref{projection0}) and (\ref{projection1}) are
``projection'' tensors that pick out certain coefficients to give the correct energy. [The
subscript $(1)$ will be explained later.]  While the contractions with the projectors might appear
complicated, their main use is to formally write the final formula \eqref{ENERGY FORMULA}.  When
we take a derivative with respect to the Riemann tensor and evaluate on the
(homogeneous) background, we always get an expression of the form
\be
\label{dldr}
 \(
\frac{\partial{\mathscr{L}}}{\partial
R^{\mu\nu}_{\rho\sigma}}\)_{\bg}=N
\delta^{\rho}_{[\mu}\delta^{\sigma}_{\nu]} ,\ee where $N$ is some
constant coefficient. The projector (\ref{projection0}) is defined
to give precisely the coefficient $N$ when acting on Eq.~\eqref{dldr}.
Thus, in practice one often simply reads off the coefficient after
computing the derivative on the background, rather than actually
performing the contraction with $P^{\rho\sigma}_{\mu\nu}$.

Similarly, for the second derivative with respect to the Riemann
tensor evaluated on the background, there are in general three terms,
\be  \label{second derivative}\(\frac{\partial^{2}\mathscr{
L}}{\partial R^{\gamma\delta}_{\alpha\beta} \partial
R^{\mu\nu}_{\rho\sigma}}\)_{\bar{g}}
=N_{1}\delta_{\mu}^{[\alpha}\delta^{\beta]}_{\nu}\delta^{\rho}_{[\gamma}\delta_{\delta]}^{\sigma}
+N_{2}\delta_{\gamma}^{[\alpha}\delta^{\beta]}_{\delta}\delta^{\rho}_{[\mu}\delta_{\nu]}^{\sigma}
+N_{3}\delta_{\delta}^{[\beta}\delta^{\alpha]}_{\mu}\delta_{\nu}^{[\sigma}\delta^{\rho]}_{\gamma}
,\ee where $N_{1},N_{2},N_{3}$ are constants.  When the projector
$P^{\phantom{(1)}\gamma\delta,\mu\nu}_
   {(1)\alpha\beta\phantom{,}\rho\sigma}$ acts on Eq.~\eqref{second derivative},
it picks out the coefficient $N_1$, but again, we can also simply
read off this coefficient by writing the expression for the second
derivative in the above form. For example, if \be \sL =
C\,R_{\alpha\beta\gamma\delta}R^{\alpha\beta\gamma\delta},
\label{Riemman2} \ee then $N_1=2C$. In other words, the second
derivative term in Eq.~\eqref{ENERGY FORMULA} roughly corresponds to
coefficients of terms with the same type of contractions as in
Eq.~(\ref{Riemman2}).

A typical example of solutions that fall off sufficiently fast at
infinity are asymptotically SdS solutions. The
asymptotic behavior of such solutions is \be h_{tt} \approx
\(\frac{r_0}{r}\)^{d-3}, \,\, \quad
h^{rr}\approx\(\frac{r_0}{r}\)^{d-3}+... \, , \label{SdS}
 \ee
where $r_0$ is a constant. For these solutions, the energy in the case of Einstein gravity is given by
\be
E_0 = \frac{(d-2) \mathrm{Vol}(S^{d-2})}{4 \kappa}r_0^{d-3},
\ee
and for a
general theory we get \be \label{energy formula}
E=\frac{(d-2)\mathrm{Vol}(S^{d-2})}{2}\,r_0^{d-3}\left[P_{\rho\sigma}^{\mu\nu}\(\frac{\partial
\mathscr{ L}}{\partial
R^{\mu\nu}_{\rho\sigma}}\)_{\bar{g}}-\frac{4\,\Lambda(d-3)}{(d-1)(d-2)}P^{\phantom{(2)}\gamma\delta,\mu\nu}_
   {(1)\alpha\beta\phantom{,}\rho\sigma}\(\frac{\partial^{2}\mathscr{
L}}{\partial R^{\gamma\delta}_{\alpha\beta} \partial
R^{\mu\nu}_{\rho\sigma}}\)_{\bar{g}}\right].
 \ee

 Another important example is the energy density of black branes in AdS. When the asymptotic behavior of the black brane solution is
\be h_{tt} \approx \frac{A}{r^{d-3}}, \,\, \quad
h^{rr}\approx\frac{A}{r^{d-3}}+... \, ,
 \ee
the energy density of the black brane is
\be
E_0 = \frac{(d-2) A}{4 \kappa}.
\ee

Let us now look at two examples.
\subsection{Example: Energy with a Gauss-Bonnet
term}

We start with the famous case of Gauss-Bonnet gravity. This term is
topological (a total derivative) in four dimensions and leads to a
ghost-free theory in any number of dimensions. The Lagrangain with
the Gauss-Bonnet term reads \be
\mathscr{L}_{GB}=\mathscr{L}_{E}+\frac{b_2}{2 \kappa} (R^2_{\mu \nu
\rho \sigma}-4 R^2_{\mu \nu}+R^2), \ee and we have \be
\frac{\partial \mathscr{L}_{GB}}{\partial
R^{\mu\nu}_{\rho\sigma}}=\frac{1}{2
\kappa}\(\delta^{\rho}_{[\mu}\delta^{\sigma}_{\nu]}+2\,b_2\,R_{\mu\nu}^{\rho\sigma}-8\,b_2\,\delta^{\rho}_{[\mu}R^{\sigma}_{\nu]}+2\,b_2\,R\,\delta^{\rho}_{[\mu}\delta^{\sigma}_{\nu]}\).
\ee
 For a homogeneous background,
 \be \label{RiemmanBgd}
\begin{array}{ccc}
\bar R_{\mu \nu}^{\rho \sigma} = \frac{4 \Lambda}{(d-1)(d-2)}
\delta_{[\mu}^\rho \delta_{\nu]}^\sigma, & \,\,\,\,\, \bar
R^{\sigma}_{\nu}=\frac{2\,\Lambda}{d-2}\delta^{\sigma}_{\nu}, &
\,\,\,\,\,
\bar R=\frac{2\,\Lambda\,d}{d-2} \,,\\
\end{array}
\ee
so
 \be \(\frac{\partial
\mathscr{L}_{GB}}{\partial
R^{\mu\nu}_{\rho\sigma}}\)_{\bg}=\frac{1}{2 \kappa}\(1+4\,\Lambda\,b_2\frac{d-3}{d-1}\)\delta^{\rho}_{[\mu}\delta^{\sigma}_{\nu]},\ee
and the projection $P_{\rho\sigma}^{\mu\nu}$ gives us the
coefficient of $\delta^{\rho}_{[\mu}\delta^{\sigma}_{\nu]}$: \be
P_{\rho\sigma}^{\mu\nu}\(\frac{\partial \mathscr{L}_{GB}}{\partial
R^{\mu\nu}_{\rho\sigma}}\)_{\bar{g}}=\frac{1}{2\kappa}\(1+4\,\Lambda\,b_2\frac{d-3}{d-1}\).
\label{GB first} \ee
The second derivative with respect to the
Riemann tensor is
 \be \frac{\partial^{2}\mathscr{ L}_{GB}}{\partial
R^{\gamma\delta}_{\alpha\beta} \partial
R^{\mu\nu}_{\rho\sigma}}=\frac{b_2}{\kappa}\(\delta_{\mu}^{[\alpha}\delta^{\beta]}_{\nu}\delta^{\rho}_{[\gamma}\delta_{\delta]}^{\sigma}
+\delta_{\gamma}^{[\alpha}\delta^{\beta]}_{\delta}\delta^{\rho}_{[\mu}\delta_{\nu]}^{\sigma}
-4\,\delta_{\delta}^{[\beta}\delta^{\alpha]}_{\mu}\delta_{\nu}^{[\sigma}\delta^{\rho]}_{\gamma}\) \,.
\ee
Since we only require the coefficient of the first type of term, namely
$\delta_{\mu}^{[\alpha}\delta^{\beta]}_{\nu}\delta^{\rho}_{[\gamma}\delta_{\delta]}^{\sigma}$, we
get
  \be
  P^{\phantom{(2)}\gamma\delta,\mu\nu}_
   {(1)\alpha\beta\phantom{,}\rho\sigma}\(\frac{\partial^{2}\mathscr{
L}_{GB}}{\partial R^{\gamma\delta}_{\alpha\beta} \partial
R^{\mu\nu}_{\rho\sigma}}\)_{\bar{g}}=
\frac{b_2}{\kappa}\,.
\label{GB second}
\ee
Substituting Eqs. (\ref{GB first}) and (\ref{GB second}) into the formula
for the energy (\ref{energy formula}), we obtain \be
E=\(1+\frac{4\,b_2\,\Lambda\,(d-4)\,(d-3)}{(d-2)(d-1)}\)\frac{(d-2) \, \mathrm{Vol}(S^{d-2})}{4 \kappa}r_0^{d-3}.
\ee
This agrees with the result in Ref.~\cite{Deser:2002jk}, which was obtained by a longer
calculation.

\subsection{Example: Energy with a six-derivative term}

Here we will consider an example with six derivatives of the metric
(cubic curvature),
 \be
\mathscr{L}_{I_1}=\mathscr{L}_{E}+\frac{c_1}{2 \kappa} R^{\mu
\nu}{}{}_{\alpha \beta} R^{\alpha \beta}{}{}_{\lambda \rho}
R^{\lambda \rho}{}{}_{\mu \nu}.
\ee
The ADT energy of this theory was previously given
in Ref.~\cite{previous}, but required a much lengthier calculation.
In this example, the first
derivative with respect to the Riemann tensor gives \be \frac{\partial
\mathscr{L}_{I_1}}{\partial
R^{\mu\nu}_{\rho\sigma}}=\frac{1}{2 \kappa}\(\delta^{\rho}_{[\mu}\delta^{\sigma}_{\nu]}+3\,c_1\,R_{\lambda\epsilon}^{\rho\sigma}R^{\lambda\epsilon}_{\mu\nu}\) \,,
\ee
and substituting the background metric leads to
 \be
\label{riemann3first}P_{\rho\sigma}^{\mu\nu}\(\frac{\partial
\mathscr{L}_{I_1}}{\partial
R^{\mu\nu}_{\rho\sigma}}\)_{\bg}=\frac{1}{2 \kappa}\(1+\frac{48\,c_1\,\Lambda^2}{(d-1)^2(d-2)^2}\).
\ee
The second derivative with respect to the Riemann tensor (which in this example is not just a constant)
is
\be
\frac{\partial^{2}\mathscr{ L}_{I_1}}{\partial
R^{\gamma\delta}_{\alpha\beta} \partial
R^{\mu\nu}_{\rho\sigma}}=\frac{1}{2 \kappa}\cdot3\,c_1\,\(R^{\alpha\beta}_{\mu\nu}\delta^{\rho}_{[\gamma}\delta^{\sigma}_{\delta]}+R^{\rho\sigma}_{\gamma\delta}\delta^{\alpha}_{[\mu}\delta^{\beta}_{\nu]}\).
\ee
When we substitute the background
metric, the second derivative becomes proportional only to the
first term in Ref.~(\ref{second derivative}),
\be
\(\frac{\partial^{2}\mathscr{ L}_{I_1}}{\partial
R^{\gamma\delta}_{\alpha\beta} \partial
R^{\mu\nu}_{\rho\sigma}}\)_{\bar{g}}=\frac{1}{2 \kappa}\cdot\frac{24\,\Lambda\,c_1}{(d-1)(d-2)}\delta^{\alpha}_{[\mu}\delta^{\beta}_{\nu]}\delta^{\rho}_{[\gamma}\delta^{\sigma}_{\delta]},
\ee
so the coefficient is just $N_1$, and $N_2 = N_3 = 0$.
Substituting this and Eq.~(\ref{riemann3first})
into the energy formula (\ref{energy formula}),
we finally obtain
 \be
E=\(1-\frac{48(2d-7)\,c_1\,\Lambda^2}{(d-2)^2(d-1)^2}\)\frac{(d-2) \, \mathrm{Vol}(S^{d-2})}{4 \kappa}r_0^{d-3},\ee
which agrees with the result in Ref.~\cite{previous}.

\section{The ADT Method}
\label{ADT} In this section we give a brief review of the ADT
method. The ADT method is similar in spirit to the Landau-Lifshitz
pseudotensor method for calculating energy \cite{LL} in
asymptotically flat curved spacetime.  In particular, one proceeds
by linearizing the equations of motion with respect to a background
spacetime.  This leads to an effective stress-energy tensor that
consists of matter sources and terms higher-order in the
perturbation.  This tensor turns out to be covariantly conserved and
can thus be used to construct a conserved charge associated with an
isometry of the background.

Let us consider some arbitrary gravitational theory with equations
of motion of the form
\begin{equation}
\label{geneom} \Phi_{\mu \nu}(g,R,\nabla R, R^2,\ldots) = \kappa
\tau_{\mu \nu},
\end{equation}
where $\kappa$ is the gravitational coupling and $\tau_{\mu \nu}$ is
the matter stress-energy tensor.  The symmetric tensor $\Phi_{\mu
\nu}$, which is the analogue of the Einstein tensor, may depend on
the metric, the curvature, derivatives of the curvature, and various
combinations thereof.  Assuming that the action is invariant under
diffeomorphisms, we obtain the geometric identity $\nabla^\mu
\Phi_{\mu \nu}=0$ (the generalized Bianchi identity) and the
covariant conservation of the stress tensor $\nabla^\mu \tau_{\mu
\nu}=0$.

Now, we further assume that there exists a background solution
$\bg_{\mu \nu}$ to the equations \eqref{geneom} with $\tau_{\mu
\nu}=0$.  Then we decompose the metric as
\begin{equation}
g_{\mu \nu}=\bg_{\mu \nu}+h_{\mu \nu} \,,
\end{equation}
where we note that the deviation $h_{\mu \nu}$ is not necessarily
infinitesimal, but it is required to fall off sufficiently fast at
infinity.   Asymptotically SdS spacetimes are a typical example meeting this requirement. By
expanding the left-hand side of Eq.~\eqref{geneom} in $h_{\mu \nu}$, the
equations of motion may be expressed as
\begin{equation}
\phi^{(1)}_{\mu \nu}= \kappa \tau_{\mu \nu}-\phi^{(2)}_{\mu
\nu}-\phi^{(3)}_{\mu \nu}\ldots\equiv \kappa T_{\mu \nu}\,,
\end{equation}
where $\phi^{(i)}_{\mu \nu}$ denotes all terms in the expansion of
$\Phi_{\mu \nu}$ involving $i$ powers of $h_{\mu \nu}$, and we have
defined the effective stress-tensor $T_{\mu \nu}$.  It then follows
from the Bianchi identity of the full theory that $\bnabla^\mu
\phi^{(1)}_{\mu \nu}=0=\bnabla^\mu T_{\mu \nu}$.

Suppose that the background spacetime admits a timelike Killing
vector $\bar{\xi}^\mu$, and let $\Sigma$ be a constant-time
hypersurface with unit normal $n^\mu$.  Then we can construct a
conserved energy in the standard way:
\begin{equation}
\label{ebulk} E=\int_\Sigma d^{d-1}x\sqrt{\bg_\Sigma}\, n_\mu T^{\mu
\nu}\bar{\xi}_\nu \,,
\end{equation}
where $\bg_\Sigma$ denotes the determinant of the induced metric on
$\Sigma$. Because $\bnabla^\mu (T_{\mu \nu} \bar \xi^\nu)=0$, it
follows that $T_{\mu \nu} \bar \xi^\nu=\bnabla^\nu \mathcal{F}_{\nu
\mu}$ for some antisymmetric tensor $\mathcal{F}_{\nu \mu}$. The
bulk integral \eqref{ebulk} can therefore be rewritten as a surface
integral over the boundary $\partial \Sigma$:
\begin{eqnarray}
\label{esurf} E=\int_{\partial\Sigma}
d^{d-2}x\sqrt{\bg_{\partial\Sigma}} \,n_\mu r_\nu \mathcal{F}^{\nu
\mu} \,,
\end{eqnarray}
where $r_\mu$ is the unit normal to the boundary.  For example, for the Einstein-Hilbert theory \eqref{Einstein action}, the explicit expression for the energy is
\begin{eqnarray}
\label{surface Einstein}
E_0 &=& \frac{1}{4 \kappa} \int_{\partial\Sigma}
d^{d-2}x\sqrt{\bg_{\partial\Sigma}} \,n_\mu r_\nu \left[\bxi_\lambda \bnabla^\mu h^{\nu \lambda}-\bxi_\lambda \bnabla^\nu h^{\mu \lambda}+\bxi^\mu \bnabla^\nu h-\bxi^\nu \bnabla^\mu h +h^{\mu \lambda} \bnabla^\nu \bxi_\lambda\right. \nonumber \\
&& \hspace{4.5cm}\left. -h^{\nu \lambda} \bnabla^\mu \bxi_\lambda +\bxi^\nu \bnabla_\lambda h^{\mu \lambda}-\bxi^\mu \bnabla_\lambda h^{\nu \lambda}+h \bnabla^\mu \bxi^\nu\right]
 \,.
\end{eqnarray}

In summary, to apply the ADT method, one linearizes the equations of
motion to obtain the stress-energy tensor, and then expresses the
conserved current $T^{\mu \nu} \bar \xi_\nu$ as a total derivative
to find the ``potential'' $\mathcal{F}^{\nu \mu}$.  Note that by
construction, the background spacetime $\bg_{\mu \nu}$ has $E=0$.

\section{The General Structure of the Stress Tensor}
\label{flat}

In Ref.~\cite{Deser:2002rt}, it was shown that the most general quadratic curvature theory has a stress tensor that is schematically of the form
\begin{eqnarray}
\label{basicT} T_{\mu \nu} &=&\alpha_1 \G^L_{\mu \nu}+\alpha_2
H^{(1)}_{\mu \nu} +\alpha_3 H^{(2)}_{\mu \nu} ,
\end{eqnarray}
where
\begin{eqnarray}
 \label{GL} \G^L_{\mu \nu} &=&
R^L_{\mu \nu}-\frac{1}{2}\bg_{\mu \nu} R_L-\frac{2
\Lambda}{d-2} h_{\mu \nu}\,,
\\
 \label{H1}
H^{(1)}_{\mu \nu} &=& \left(\bg_{\mu \nu} \bar{\Box} -\bn_{\mu}
\bn_{\nu} +\frac{2 \Lambda}{d-2} \bg_{\mu \nu} \right)R_L
\\
 \label{H2}
H^{(2)}_{\mu \nu} &=&\bar{\Box} \G^L_{\mu \nu}-\frac{2
\Lambda}{d-2}\bg_{\mu \nu} R_L
\end{eqnarray}
and the $\alpha_i$ are constants. Here $R^L_{\mu \nu}$ is the
linearized Ricci tensor \be R^L_{\mu
\nu}=R_{\mu\nu}-\bar{R}_{\mu\nu}=\frac{1}{2}\(-\bar \Box
h_{\mu\nu}-\bn_{\mu}\bn_{\nu}h+\bn^{\sigma}\bn{\nu}h_{\sigma\mu}+\bn^{\sigma}\bn_{\mu}h_{\sigma\nu}\)\,
, \ee and $R_{L}$ is the linearized Ricci scalar \be
R_{L}=\bn^{\sigma}\bn^{\mu}h_{\sigma\mu}-\bar \Box h
-\frac{2\,\Lambda}{d-2}h\, .\ee   Note that the tensors $\G^L_{\mu
\nu}, H^{(i)}_{\mu \nu}$ are \emph{each} divergence free:
 \be
 \bn^{\mu} \G^L_{\mu \nu}=\bn^{\mu} H^{(1)}_{\mu \nu}= \bn^{\mu} H^{(2)}_{\mu \nu}= 0 \,.
 \ee
It was later found in Ref.~\cite{previous} that the stress tensor for a certain cubic curvature theory took exactly the same form (the only modification was to the values of the coefficients $\alpha_i$) and it was suggested that this observation might hold more generally\footnote{This general claim was also stated in Ref.~\cite{Tekin}.}.  In this section, we will argue that this is indeed the case for any theory of the form \eqref{general lagrangian}.

The basic idea is as follows.  We saw in the previous section that
the ADT stress tensor is given by the linearized (in $h$) equations
of motion.  This means that $T_{\mu \nu}$ only depends on the action
to $O(h^2)$.  Now, expanding the general action \eqref{general
lagrangian} to $O(h^2)$ involves expanding the Riemann tensor, which
contains terms of the form $\bnabla \bnabla h$.   Hence, this can
yield terms of \emph{at most} four derivatives. This suggests quite
generally that the basic form of the stress tensor does not change from
Eq.~\eqref{basicT} even if the action contains more than two powers of
the Riemann tensor\footnote{For a flat background, it is clear that
no term with more than two powers of the Riemann tensor can
contribute to the $O(h^2)$ part of the action, since $\bar R_{\mu
\nu \rho \sigma} = 0$.}. We will show that there exists a basis of three different components for the
stress-energy tensor [to $O(h^2)$ and up to four derivatives],
and that this basis can be chosen to correspond to $\G^L_{\mu \nu}$,
$H^{(1)}_{\mu \nu}$, and $H^{(2)}_{\mu \nu}$.

To demonstrate this claim in more detail, we first consider the most general $O(h^2)$ action with two derivatives:
\be
I_2=\int d^{d}x \(\beta_1 \p^{\rho}h^{\mu\nu}\p_{\rho}h_{\mu\nu}+\beta_2\p^{\mu}h^{\nu\rho}\p_{\nu}h_{\mu\rho}+\beta_3\p^{\mu}h
\p^{\lambda}h_{\lambda\mu}+\beta_4\p^{\mu}h\p_{\mu}h\) \,.
\ee
 For simplicity we work in the case of a flat background, $\bg_{\mu \nu} = \eta_{\mu \nu}$.  The generalization to a curved background will be discussed later.  Varying $I_2$ with respect to $h_{\mu \nu}$ yields the stress tensor
\be \label{stress2} T_{\alpha\beta}=-\frac{\delta I_2}{\delta
h^{\alpha\beta}}= 2\beta_1 \bar \Box h^{\alpha\beta}+2 \beta_2
\p_{\nu}\p^{(\alpha}h^{\beta\nu)} +\beta_3
\,\eta^{\alpha\beta}\p^{\mu}\p^{\lambda}h_{\lambda\mu}+\beta_3
\p^{\alpha}\p^{\beta}h+2\beta_4 \eta^{\alpha\beta} \bar\Box h . \ee
If we impose conservation of the stress tensor, we obtain \be 0 =
\p_{\alpha}T^{\alpha\beta}=(2 \beta_1+\beta_2)\bar
\Box\p_{\alpha}h^{\alpha\beta}+(\beta_2+\beta_3)\p^{\beta}\p_{\nu}\p_{\alpha}h^{\nu\alpha}+(\beta_3+2
\beta_4) \bar \Box\p^{\beta} h \,,
 \ee
so equating the coefficients to zero gives
\bea
\beta_2=-2 \beta_1\,, \quad \beta_3 =2 \beta_1\,, \quad \beta_4=- \beta_1 \,.
\eea
Substituting these relations into Eq.~\eqref{stress2} gives $T_{\mu \nu} = -4 \beta_1 \G^L_{\mu \nu}$.  It follows that the Lagrangian
 \be
L_\G=
\frac{1}{4}\left(2\p^{\mu}h^{\nu\rho}\p_{\nu}h_{\mu\rho}+\p^{\mu}h\p_{\mu}h-2\p^{\mu}h\p^{\lambda}h_{\lambda\mu}-\p^{\rho}h^{\mu\nu}\p_{\rho}h_{\mu\nu}
\right) \ee yields a conserved stress tensor that is precisely
$\G^L_{\mu \nu}$ (in a flat background). Note that this is also just
the famous Fierz-Pauli Lagrangian~\cite{fierz-pauli}.

Now let us repeat the same procedure for the most general $O(h^2)$
action with four derivatives: \be I_4=\int d^{d}x \(\beta_5
\p_{\alpha}\p_{\beta}h^{\alpha\beta}\p_{\gamma}\p_{\delta}h^{\gamma\delta}+\beta_6
\bar
\Box\,h^{\beta\gamma}\p_{\alpha}\p_{\gamma}h^{\alpha}_{\beta}+\beta_7
\bar\Box h^{\alpha\beta}\p_{\alpha}\p_{\beta}h+\beta_8 \, \bar
\Box\,h_{\beta\gamma} \bar \Box\,h^{\beta\gamma}+\beta_9 \bar \Box h
\bar \Box h\)\,. \ee The corresponding stress tensor is \bea
\label{stress4} T_{\mu\nu} &=&-\frac{\delta I_4}{\delta h^{\mu\nu}}=
-(2
\beta_5\p_{\mu}\p_{\nu}\p_{\gamma}\p_{\delta}h^{\gamma\delta}+2\beta_6
\bar \Box\p_{\alpha}\p_{(\mu}h_{\nu)}^{\alpha}+ \beta_7 \( \bar
\Box\p_{\mu}\p_{\nu}h +\eta_{\mu\nu} \bar
\Box\p_{\alpha}\p_{\beta}h^{\alpha\beta}\) \nonumber\\&&\qquad
\qquad+2\beta_8 \bar \Box^2 h_{\mu\nu}+2\beta_9 \,\eta_{\mu\nu}\bar
\Box^2h) \eea and imposing $\p^{\mu}T_{\mu\nu} = 0$ gives
 \bea
 \beta_7=-\beta_6-2 \beta_5 \,, \quad
 \beta_8=-\frac{\beta_6}{2}\,, \quad
 \beta_9=\beta_5+\frac{\beta_6}{2} \,.
 \eea
We see that there are two independently conserved tensors, and
substituting these relations into Eq.~\eqref{stress4} gives $T_{\mu \nu}
= -2 \beta_5 H^{(1)}_{\mu \nu}+ 2 \beta_6 H^{(2)}_{\mu \nu}$.  It
follows that the Lagrangian
\begin{equation}
L_{H^{(1)}}=\frac{1}{2}\left(\p_{\alpha}\p_{\beta}h^{\alpha\beta}\p_{\gamma}\p_{\delta}h^{\gamma\delta}-2 \bar \Box
h^{\alpha\beta}\p_{\alpha}\p_{\beta}h+\bar \Box
h \bar \Box h \right)
\end{equation}
yields a conserved stress tensor that is precisely $H^{(1)}_{\mu \nu}$ and the Lagrangian
\begin{equation}
L_{H^{(2)}} = -\frac{1}{4} \left(2 \bar \Box\,h^{\beta\gamma}\p_{\alpha}\p_{\gamma}h^{\alpha}_{\beta}-2 \bar \Box
h^{\alpha\beta}\p_{\alpha}\p_{\beta}h- \bar \Box\,h_{\beta\gamma} \bar \Box\,h^{\beta\gamma}+ \bar \Box
h \bar \Box h \right)
 \end{equation}
 yields a conserved stress tensor that is precisely $H^{(2)}_{\mu
 \nu}$. Thus, we see that, in a flat background, there are at most two
 possible conserved combinations of terms with four derivatives in
 the Lagrangian.

 The above calculation can in principle be repeated for the case of a curved background spacetime, but it becomes complicated by the fact that the covariant derivatives no longer commute.  The key point, however, is that commuting two derivatives in a given expression only produces extra terms of lower differential order. Thus, the expressions for $L_{H^{(i)}}$ now must include two-derivative and zero-derivative terms,
 but the highest derivative order (four) terms are the same as in a flat background.
 Once these are fixed, the two-derivative and zero-derivative terms are chosen by requiring
 that each $H^{(i)}_{\mu \nu}$ is separately conserved.  A similar
argument shows that the unique conserved quantity consisting of
two-derivative and 0-derivative terms is $\G^L_{\mu \nu}$.
In other words, the terms with the highest derivatives (four derivatives
in $H^{(i)}_{\mu \nu}$ and two derivatives in $\G^L_{\mu \nu}$) are
the same for curved and flat backgrounds.

 In summary, we have just seen that any $O(h^2)$ action with no more than four derivatives produces a conserved stress tensor with the same structure as Eq.~\eqref{basicT}.  Combining this with the argument that any theory
 $\sL = \sL(R^{\mu \nu}_{\rho \sigma},g_{\mu\nu})$ expanded to $O(h^2)$ cannot have terms with more than four derivatives, we conclude that any such theory also has a stress tensor of the form \eqref{basicT}.

 \section{The General Formula for the Stress Tensor (Flat Background)}
\label{flat2}
 In this section we derive an efficient method to extract the coefficients $\alpha_i$ in the stress tensor \eqref{basicT}
 given a Lagrangian $\sL = \sL(R^{\mu \nu}_{\rho \sigma},g_{\mu\nu})$ for a flat
 background. The generalization for a curved background will be done
 in the next section.

 We wish to expand the action $\sqrt{-g} \, \sL$ to second order in the perturbation $h$.  The Lagrangian can be expanded as
 \begin{equation}
\delta \mathscr{L} = \left(\frac{\p \mathscr{L}}{\partial R^{\mu
\nu}_{\rho \sigma}}\right)_{\bg}\, \delta R^{\mu \nu}_{\rho \sigma}+\frac{1}{2}\,\left(\frac{\p^2
\mathscr{L}}{\p R^{\mu \nu}_{\rho \sigma} \, \p R_{\alpha
\beta}^{\gamma \delta}} \right)_{\bg}\,\delta R^{\mu \nu}_{\rho \sigma} \delta R_{\alpha \beta}^{\gamma \delta}+O(h^3) \,.
\end{equation}
The variation of the Riemann tensor is
\begin{equation}
\label{Rvariation2} \delta
R^\rho{}_{\mu\lambda\nu}= R^\rho{}_{\mu\lambda\nu}-\bar R^\rho{}_{\mu\lambda\nu}=\bnabla_\lambda \delta \Gamma^{\rho}_{\nu \mu} - \bnabla_\nu \delta \Gamma^\rho_{\lambda \mu}+ \delta \Gamma^\rho_{\lambda \delta} \delta \Gamma^{\delta}_{\mu \nu}- \delta \Gamma^\rho_{\delta \nu} \delta \Gamma^\delta_{\mu \lambda}
\end{equation}
where
\begin{eqnarray}
\delta \Gamma^\rho_{\mu \nu} &\equiv& \frac{1}{2}g^{\rho \kappa}  \left(\bnabla g_{\nu \kappa} +\bnabla g_{\mu \kappa} - \bnabla_\kappa g_{\mu \nu} \right) \\
&=&  \Upsilon_{\mu \nu}{}^\rho -h^{\rho \kappa} \Upsilon_{\mu \nu \kappa} + O(h^3)
\end{eqnarray}
and
\begin{equation}
\Upsilon_{\alpha\beta\gamma}=
\frac{1}{2}\left(\overline{\nabla}_{\alpha}h_{\beta\gamma} +
\overline{\nabla}_{\beta}h_{\alpha\gamma}
-\overline{\nabla}_{\gamma}h_{\alpha\beta}\right).
\end{equation}
By convention, indices of $\Upsilon_{\alpha \beta \gamma}$ are raised/lowered with the background metric or its inverse.  Note that each factor of $\delta R^{\mu \nu}_{\rho \sigma}$ contributes \emph{at least} one $h_{\mu \nu}$ and two derivatives.

In the remainder of this section we restrict to the case of a flat background, $\bg_{\mu \nu} = \eta_{\mu \nu}$.  Now, the terms in the action with two derivatives and two $h$'s can only arise from expanding the term with one $\delta R^{\mu \nu}_{\rho \sigma}$, that is
\begin{equation}
\label{2h2d}
\left(\frac{\p \mathscr{L}}{\partial R^{\mu
\nu}_{\rho \sigma}}\right)_{\bg}\, \sqrt{-g} \, \delta R^{\mu \nu}_{\rho \sigma} \,.
\end{equation}
Since $\sL$ is a function only of $g_{\mu \nu}$ and $R^{\mu \nu}_{\rho \sigma}$, it follows that $\partial
\mathscr{L}/\partial R^{\mu \nu}_{\rho \sigma}$
evaluated on a homogeneous background can only be a function of
$\bg_{\mu \nu}$. Furthermore, this quantity has the same symmetries
as the Riemann tensor, so it must take the general form
\begin{equation}
\label{dLdR}
\left(\frac{\partial \mathscr{L}}{\partial R^{\mu \nu}_{\rho
\sigma}}\right)_{\bg}  = N \delta^{[\rho}_\mu \delta^{\sigma]}_\nu
\end{equation}
for some constant $N$.  Formally, this constant can be expressed as a ``projection''
\begin{equation}
N = P_{\rho\sigma}^{\mu\nu} \left(\frac{\partial \mathscr{L}}{\partial R^{\mu \nu}_{\rho
\sigma}}\right)_{\bg}
\end{equation}
for the projection tensor
\begin{equation}
P_{\rho\sigma}^{\mu\nu}=\frac{2\,\delta_{[\rho}^\mu \delta_{\sigma]}^\nu}{d(d-1)} \,.
\end{equation}
Inserting Eq.~\eqref{dLdR} into Eq.~\eqref{2h2d}, we see that we simply need
the expansion of $N \sqrt{-g} \,R$.  This is of course just the
Einstein-Hilbert action (up to the overall factor $N$), whose
expansion is well-known to give the Fierz-Pauli action (see,
e.g.,Ref.~\cite{ortin}), $N L_\G$.

The $O(h^2)$ terms in the action with four derivatives can only arise from the term
\begin{equation}
\label{L0R1R1}
\sqrt{-\bg} \,\left(\frac{\p^2
\mathscr{L}}{\p R^{\mu \nu}_{\rho \sigma} \, \p R_{\alpha
\beta}^{\gamma \delta}} \right)_{\bg} \,\delta R^{\mu \nu}_{\rho \sigma} \delta R_{\alpha \beta}^{\gamma \delta} \,,
\end{equation}
and for this we just need the linear term in $\delta R^{\mu \nu}_{\rho \sigma}$,
\begin{equation}
\label{RiemannLinear} \left(\delta R^{\rho \sigma}_{\lambda
\nu}\right)^{L} =   \p_\lambda \Upsilon_{\nu}{}^ \sigma{}^{\rho} -
\p_\nu \Upsilon_{\lambda}{}^{\sigma \rho} \,.
\end{equation}
Now, the second derivative evaluated on a homogeneous background can only be a function of $\bg_{\mu \nu}$, and this quantity must have the same index symmetries as the product of two Riemann tensors.  Hence, there can be three independent contributions
\be
\label{d2ldrdr}
\(\frac{\partial^{2}\mathscr{ L}}{\partial
R^{\gamma\delta}_{\alpha\beta} \partial
R^{\mu\nu}_{\rho\sigma}}\)_{\bar{g}}
=N_{1}\delta_{\mu}^{[\alpha}\delta^{\beta]}_{\nu}\delta^{\rho}_{[\gamma}\delta_{\delta]}^{\sigma}
+N_{2}\delta_{\gamma}^{[\alpha}\delta^{\beta]}_{\delta}\delta^{\rho}_{[\mu}\delta_{\nu]}^{\sigma}
+N_{3}\delta_{\delta}^{[\beta}\delta^{\alpha]}_{\mu}\delta_{\nu}^{[\sigma}\delta^{\rho]}_{\gamma}
\ee
for some constants $N_i$.  This is similar to the statement that there are only three independent curvature invariants formed from contracting two Riemann tensors.  Formally, these constants can be expressed by acting with projectors
\begin{equation}
N_i = P^{\phantom{(2)}\gamma\delta,\mu\nu}_
   {(i)\alpha\beta\phantom{,}\rho\sigma}\(\frac{\partial^{2}\mathscr{
L}}{\partial R^{\gamma\delta}_{\alpha\beta} \partial
R^{\mu\nu}_{\rho\sigma}}\)_{\bar{g}} \,,
\end{equation}
where
\begin{equation}
P^{\phantom{(1)}\gamma\delta,\mu\nu}_
   {(i)\alpha\beta\phantom{,}\rho\sigma}=
a_i\delta^{\mu}_{[\alpha } \delta^{\nu}_{\beta]}
\delta_{\rho}^{[\gamma } \delta_{\sigma}^{\delta]}
+b_i \delta^{\gamma}_{[\alpha } \delta^{\delta}_{\beta]}
\delta_{\rho}^{[\mu } \delta_{\sigma}^{\nu]}
+c_i \delta^{\delta}_{[\beta } \delta^{\mu}_{\alpha]}
\delta_{[\sigma}^{\nu} \delta_{\rho]}^{\gamma} \,.
\end{equation}
The coefficients are
\bea
 a_1&=&b_2=(d-1)^3\, p \phantom{-(d-2)} \, \\
 a_2&=&b_1=-(d-1)\,p  \\
a_3&=&b_3=c_1=c_2=-(d-2)(d-1)\,p  \\
 c_3&=&(d^2-d+2)(d-2)\,p \,,
\eea
with
\be
p\equiv\frac{4}{d(d^2-1)(d-1)(d-2)(d^2-2\,d-2)} \,.
\ee
The next step is to insert Eq.~\eqref{d2ldrdr} into Eq.~\eqref{L0R1R1} and use Eq.~\eqref{RiemannLinear}.  We treat the three contractions separately.  The first is analogous to $R_{\mu \nu \rho \sigma}^2$ and gives
\begin{equation}
N_{1}\delta_{\mu}^{[\alpha}\delta^{\beta]}_{\nu}\delta^{\rho}_{[\gamma}\delta_{\delta]}^{\sigma}
\left(\delta R^{\mu \nu}_{\rho \sigma}\right)^{L} \left(\delta
R_{\alpha \beta}^{\gamma \delta} \right)^{L}= 2 N_1 L_{H^{(1)}}+4
N_1 L_{H^{(2)}} \,.
\end{equation}
The second contraction is analogous to $R^2$ and gives
\begin{equation}
N_{2}\delta_{\gamma}^{[\alpha}\delta^{\beta]}_{\delta}\delta^{\rho}_{[\mu}\delta_{\nu]}^{\sigma}
\left(\delta R^{\mu \nu}_{\rho \sigma}\right)^{L} \left(\delta
R_{\alpha \beta}^{\gamma \delta} \right)^{L}= 2 N_2 L_{H^{(1)}}
\,.\end{equation} The last contraction is analogous to $R_{\mu
\nu}^2$ and gives
\begin{equation}
N_{3}\delta_{\delta}^{[\beta}\delta^{\alpha]}_{\mu}\delta_{\nu}^{[\sigma}\delta^{\rho]}_{\gamma}
\left(\delta R^{\mu \nu}_{\rho \sigma}\right)^{L} \left(\delta
R_{\alpha \beta}^{\gamma \delta} \right)^{L}= N_3 L_{H^{(1)}}+N_3
L_{H^{(2)}} \,.
\end{equation}
Thus the relevant part of the expanded action is
\begin{equation}
\delta (\sqrt{-g} \,\sL) = N L_\G + \left(N_1 + N_2 +\frac{N_3}{2} \right) L_{H^{(1)}}+ 2\left(N_1+\frac{N_3}{4}\right) L_{H^{(2)}}+\ldots
\end{equation}
and the corresponding stress tensor is
\begin{eqnarray}
T_{\mu \nu} &=& N \G^L_{\mu \nu} +\left(N_1+N_2 +\frac{1}{2} N_3\right) H^{(1)}_{\mu \nu}+\left(2 N_1  +\frac{1}{2} N_3\right) H^{(2)}_{\mu \nu} \,.
\end{eqnarray}

\section{The General Formula for the Stress Tensor (Curved Background)}
\label{energy-momentum}

The procedure described previously for a flat background should in
principle generalize to a curved background.  The calculation
becomes cumbersome, however, since the covariant derivatives no
longer commute.  Instead, we shall adopt a different approach that
turns out to be much more straightforward.

It was argued
in Refs.~\cite{Tekin,Hindawi:1995cu,Gullu:2010em,Gullu:2010vw,Sisman:2012rc}
that any higher-curvature theory which is polynomial in the Riemann
tensor and its contractions can be reduced to an ``effective
quadratic curvature'' action with the same propagator.  Since the
propagator also only depends on the action up to order $h^2$, we can
adapt this procedure to determine the ADT stress-tensor for a
general theory.

Consider expanding the Lagrangian of a generic higher-curvature
theory of the form $\mathscr{L} = \mathscr{L}(R^{\mu \nu}_{\rho
\sigma})$,
\begin{equation}
\mathscr{L} = \mathscr{L}(\bar R^{\mu \nu}_{\rho
\sigma})+\left(\frac{\partial \mathscr{L}}{\partial R^{\mu
\nu}_{\rho \sigma}}\right)_{\bg} (R^{\mu \nu}_{\rho \sigma}-\bar
R^{\mu \nu}_{\rho \sigma})+\frac{1}{2}\,\left(\frac{\partial^2
\mathscr{L}}{\partial R^{\mu \nu}_{\rho \sigma} \partial R_{\alpha
\beta}^{\gamma \delta}}\right)_{\bg} (R^{\mu \nu}_{\rho \sigma}-\bar
R^{\mu \nu}_{\rho \sigma}) (R_{\alpha \beta}^{\gamma \delta}-\bar
R_{\alpha \beta}^{\gamma \delta})+\ldots
\end{equation}
Here the dots represent terms which are necessarily of order $h^3$
and therefore are not relevant to the ADT energy.  Next we substitute
the general expressions for the derivatives of the Lagrangian with
respect to the Riemann tensor evaluated on the background, which were previously given in Eqs.~\eqref{dLdR} and \eqref{d2ldrdr}.
 Using Eq. (\ref{RiemmanBgd}) and collecting coefficients of the full Riemann tensor terms, we
obtain the effective quadratic theory
\begin{equation}
\mathscr{L}_{eff} = \frac{1}{2 \tilde \kappa} (R - 2 \Lambda^{eff}_0) +\alpha
R^2+\beta R_{\mu \nu}^2 +\gamma (R_{\mu \nu \rho \sigma}^2 - 4
R_{\mu \nu}^2 +R^2) \,.
\end{equation}
Here we have defined
\begin{eqnarray}
\frac{1}{2\tilde \kappa} &\equiv& N - \frac{2 \Lambda d}{d-2} N_2 -  \frac{4 \Lambda}{(d-1)(d-2)} N_1- \frac{2 \Lambda}{d-2} N_3 \\
\alpha &\equiv&  \frac{1}{2} (N_2 - N_1) \\
\beta &\equiv&  \frac{1}{2} (N_3 + 4 N_1)\\
\gamma &\equiv&  \frac{1}{2} N_1
\end{eqnarray}
and the ``bare'' cosmological constant for the effective theory is
\begin{equation}
\Lambda_0^{eff} = - \tilde \kappa \left(\mathscr{L}(\bar R^{\mu
\nu}_{\rho \sigma}) - \frac{2 \Lambda d}{d-2} N+\frac{2 \Lambda^2
d^2}{(d-2)^2} N_2+\frac{4 \Lambda^2 d}{(d-2)^2(d-1)} N_1 + \frac{2 d
\Lambda^2}{(d-2)^2} N_3 \right) \,.
\end{equation}
Now, the most general quadratic curvature theory has already been
treated in Ref.~\cite{Deser:2002jk}.  The result is that the stress
tensor is
\begin{eqnarray}
\label{quadT}
T_{\mu \nu} = \left( \frac{1}{2\tilde \kappa}+ \frac{4 d \Lambda}{d-2} \alpha+\frac{4 \Lambda}{d-1} \beta +\frac{4(d-3)(d-4) \Lambda}{(d-2)(d-1)} \gamma  \right) \G^L_{\mu \nu} +\left(2 \alpha +\beta \right) H^{(1)}_{\mu \nu} +\beta
H^{(2)}_{\mu \nu} \,,
\end{eqnarray}
where
 $\G^{L}_{\mu\nu}$, $H^{(1)}_{\mu\nu}$, and $H^{(2)}_{\mu\nu}$
 were given in Eqs.~(\ref{GL})-(\ref{H2}).
Furthermore, the effective cosmological constant $\Lambda$ is fixed
by evaluating the equation of motion on the background solution:
\begin{equation}
\left[\frac{(d-4)(d \alpha
+\beta)}{(d-2)^2}+\frac{(d-3)(d-4)\gamma}{(d-1)(d-2)} \right]
\Lambda^2+ \frac{\Lambda- \Lambda_0^{eff}}{4 \tilde \kappa} = 0 \,.
\end{equation}
Substituting the above expressions for $\tilde \kappa, \alpha, \beta, \gamma$ into Eq.
\eqref{quadT} yields
\begin{eqnarray}
\label{Tfinal}
T_{\mu \nu} = \left[N-\frac{4 \Lambda}{d-2} N_1 -\frac{2 \Lambda}{(d-1)(d-2)} N_3\right] \G^L_{\mu \nu} +\left[N_1+N_2 +\frac{1}{2} N_3\right] H^{(1)}_{\mu \nu} +\left[2 N_1  +\frac{1}{2} N_3\right] H^{(2)}_{\mu \nu}\,.
\end{eqnarray}
Note that for $\Lambda = 0$, this agrees with the result of the previous section.

\subsection{Examples}

Let us look at some examples for the formula (\ref{Tfinal}) for the
stress-energy tensor. Let us start with the simple theory
\begin{equation}
\mathscr{L}=  R_{\mu \nu \rho \sigma}^2 = R^{\mu \nu}_{\rho \sigma}
R_{\mu \nu}^{\rho \sigma}.
\end{equation}
The coefficients $N,N_i$ are computed as described previously by
taking derivatives with respect to the Riemann tensor and evaluating
on the background AdS solution.  We find that
\begin{eqnarray}
N &=& P_{\lambda \kappa}^{ \eta \varepsilon} \left(\frac{\partial \mathscr{L}}{\partial R_{\lambda \kappa}^{ \eta \varepsilon}}\right)_{\bg} = P_{\lambda \kappa}^{ \eta \varepsilon} \left( 2 R^{\lambda \kappa}_{ \eta \varepsilon}  \right)_{\bg} = P_{\lambda \kappa}^{ \eta \varepsilon}  \left( \frac{4 \Lambda}{(d-1)(d-2)} (\delta^\lambda_\eta \delta^\kappa_\varepsilon - \delta^\lambda_\varepsilon \delta^\kappa_\eta)  \right) \nonumber \\
&=& \frac{8 \Lambda}{(d-1)(d-2)}
\end{eqnarray}
and
\begin{eqnarray}
N_i = P^{\phantom{(i)}\eta \varepsilon,
\alpha \beta}_{(i)\lambda \kappa \phantom{,}\gamma \delta}  \left(\frac{\partial^2 \mathscr{L}}{\partial R_{\alpha
\beta}^{\gamma \delta} \partial R_{\lambda \kappa}^{ \eta
\varepsilon} }\right)_{\bg} =P^{\phantom{(i)}\eta \varepsilon,
\alpha \beta}_{(i)\lambda \kappa \phantom{,}\gamma \delta}  \left(
\frac{\partial}{\partial R_{\alpha \beta}^{\gamma \delta}}2
R^{\lambda \kappa}_{ \eta \varepsilon}  \right)_{\bg} = 2
P^{\phantom{(i)}\eta \varepsilon,
\alpha \beta}_{(i)\lambda \kappa \phantom{,}\gamma \delta}   \, \delta^\lambda_\gamma \delta^\kappa_\delta
\delta^\alpha_\eta \delta^\beta_\varepsilon \,,
\end{eqnarray}
so $N_1 = 2$ , $N_2 =N_3= 0$.  Using Eq.~\eqref{Tfinal}, we obtain
\begin{equation}
T_{\mu \nu} = -\frac{8 \Lambda}{d-1} \G^L_{\mu \nu} + 2 H^{(1)}_{\mu
\nu} + 4 H^{(2)}_{\mu \nu} \,,
\end{equation}
which matches the result of Ref.~\cite{Deser:2002jk}.

For a more complicated example, consider the six-derivative theory
\begin{eqnarray}
\label{ActionEg} \mathscr{L} &=& R+b_1 R^2 +b_2 (R^2_{\mu \nu \rho
\sigma}-4 R^2_{\mu \nu}+R^2) +b_3 R^2_{\mu \nu}  +c_1 R^{\mu
\nu}{}{}_{\alpha \beta} R^{\alpha \beta}{}{}_{\lambda \rho}
R^{\lambda \rho}{}{}_{\mu
\nu}\nonumber \\
&&  \qquad+c_2 R^{\mu \nu}{}{}_{\rho \sigma} R^{\rho
\tau}{}{}_{\lambda \mu} R^{\sigma}{}_{\tau}{}^\lambda{}_\nu \,,
\end{eqnarray}
whose stress tensor was computed explicitly in Ref.~\cite{previous}.
The results are summarized in the following table:
\begin{table}[h]
\vspace{.5cm}
\begin{tabular}{|c|c|c|c|c|}
\hline
$\mathscr{L}$ & \quad $N$ \quad & \, $N_1$ \quad & $N_2$ \quad & $N_3$ \\
\hline \hline
$R$ & 1 & 0 & 0 & 0  \\
\hline
\,$R^2$\, & $2 d (d-1) k$ & 0 & 2 & 0 \\
\hline
\,$R_{\mu \nu}^2$\, & $2 (d-1) k$ & 0 & 0 & 2 \\
\hline
\,$R_{\mu \nu \rho \sigma}^2$\, & $4 k$ & 2 & 0 & 0 \\
\hline
\,\,$R_{\mu \nu \rho \sigma}^2-4 R_{\mu \nu}^2 +R^2$\,\, & \, $2(d-3)(d-2) k \,\,$ & 2 & 2 & $-8$ \,\\
\hline \,$R^{\mu \nu}{}{}_{\alpha \beta} R^{\alpha
\beta}{}{}_{\lambda \rho} R^{\lambda \rho}{}{}_{\mu
\nu}$\, & $12 k^2$ & $12 k$ & 0 & 0 \\
\hline $R^{\mu \nu}{}{}_{\rho \sigma} R^{\rho \tau}{}{}_{\lambda
\mu} R^{\sigma}{}_{\tau}{}^\lambda{}_\nu$ & $3(d-2) k^2$ & $-3 k$ \, & 0 & \,$6 k$ \, \\
\hline
\end{tabular}
\vspace{.2cm}
\end{table}

where
\begin{equation}
k \equiv \frac{2 \Lambda}{(d-1)(d-2)} \,.
\end{equation}
Substituting the above results into Eq.~\eqref{Tfinal} gives
\begin{equation}
T_{\mu \nu} = \alpha_1 \G^L_{\mu \nu} + \alpha_2 H^{(1)}_{\mu \nu} +
\alpha_3 H^{(2)}_{\mu \nu} \,,
\end{equation}
where
\begin{eqnarray}
\alpha_1 &=& 1+\frac{4 d \Lambda  b_1}{d-2}+\frac{4  (d-3)(d-4) \Lambda b_2}{(d-2)(d-1)}+\frac{4 \Lambda b_3}{d-1}-\frac{48 (2d-3)\Lambda^2 c_1}{(d-2)^2(d-1)^2}+\frac{36\Lambda^2 c_2}{(d-2)(d-1)^2} \nonumber \\
\alpha_2 &=& 2 b_1 +b_3 +\frac{24 \Lambda c_1}{(d-2)(d-1)} \\
\alpha_3 &=& b_3+\frac{48 \Lambda c_1}{(d-2)(d-1)}-\frac{6 \Lambda
c_2}{(d-2)(d-1)}\,. \nonumber
\end{eqnarray}
 This reproduces precisely the stress tensor obtained in Ref.~\cite{previous}.

\section{The Derivation of the Energy Formula}
\label{the derivation}
 Given the result \eqref{Tfinal}, the final
step in the derivation of the energy formula is to write
$\bar{\xi}_{\nu}T^{\mu\nu}$ as a total derivative.   For this
purpose, we follow the steps in Ref.~\cite{Deser:2002rt}. For
the first term, $\bar{\xi}_{\nu} \G_L^{\mu\nu}$, the result has already been
given in Eq.~\eqref{surface Einstein}.
 It is straightforward to show that the second term
can be written as
 \be \bar{\xi}_{\nu}\,H^{(1)\,\mu \nu}
=\bn_{\alpha}\(\bar{\xi}^{\mu}\bn^{\alpha}
R_L+R_L\bn^{\mu}\bxi^{\alpha}-\bxi^{\alpha}\bn^{\mu}R_L\) \,. \ee
The third term, $\bar{\xi}_{\nu}\,H^{(2)\,\mu \nu}$, is more
complicated and turns out to give an additional contribution of the
form $\bar{\xi}_{\nu}\G_L^{\mu\nu}$. To see this, we can rewrite
\bea \label{boxG}
 \bxi_{\nu}\bar \Box \G^{
\mu\nu}_{L}=\bn_{\alpha}\(\bxi_{\nu}\bn^{\alpha}\G_L^{\mu\nu}-\bxi_{\nu}\bn^{\mu}\G_L^{\alpha\nu}-\G_L^{\mu\nu}\bn^{\alpha}\bxi_{\nu}+\G_L^{\alpha\nu}\bn^{\mu}\bxi_{\nu}\)
+
\G_L^{\mu\nu}\bar \Box\bxi_{\nu}+\bxi_{\nu}\bn_{\alpha}\bn^{\mu}\G_L^{\alpha\nu}-\G_L^{\alpha\nu}\bn_{\alpha}\bn^{\mu}\bxi_{\nu} \,.
\eea
Since $\bxi^{\nu}$ is a Killing vector, it satisfies
 \be
\bn_{\alpha}\bn_{\mu}\bxi_{\nu}=\bar{R}^{\rho}{}_{\nu\mu\alpha}\bxi_{\rho}=\frac{2\Lambda}{(d-2)(d-1)}\(\bg_{\nu\alpha}\bxi_{\mu}-\bg_{\alpha\mu}\bxi_{\nu}\)
\ee
 \be
  \bar \Box\bxi_{\nu}=-\frac{2\Lambda}{d-2}\bxi_{\nu}
 \ee
 Then the last terms of Eq. \eqref{boxG} simplify to
 \be
\bxi_{\nu}\bn_{\alpha}\bn^{\mu}\G_{L}^{\alpha\nu}=
\frac{2\,\Lambda\,d}{(d-2)(d-1)}\bxi_{\nu}\G_{L}^{\mu\nu}+\frac{\Lambda}{d-1}\bn^{\mu}R_{L}
\ee
\bea
\G_L^{\mu\nu}\Box\bxi_{\nu}+\bxi_{\nu}\bn_{\alpha}\bn^{\mu}\G_L^{\alpha\nu}-\G_L^{\alpha\nu}\bn_{\alpha}\bn^{\mu}\bxi_{\nu}
=\frac{\Lambda}{d-1}\(\frac{4}{d-2}\,\G^{\mu\nu}_{L}\bxi_{\nu}+\bxi^{\mu}R_{L}-\frac{2}{d-2}\G_{L}\bxi^{\mu}\) \,.
\eea

Using these results, we find that the final form of the conserved energy is
\bea
\label{g charge}
E &=&\(N-\frac{4\,\Lambda(d-3)}{(d-1)(d-2)}N_1\)2\kappa E_0
+\(N_1+N_2+\frac{N_3}{2}\)\int_{\partial\Sigma}
d^{d-2}x\sqrt{\bg_{\partial\Sigma}} \,n_\mu r_\nu\(\bar{\xi}^{\mu}\bn^{\nu}
R_L+R_L\bn^{\mu}\bxi^{\nu}-\bxi^{\nu}\bn^{\mu}R_L\) \nonumber\\
&&\quad +\(2\,N_{1}+\frac{N_{3}}{2}\)\int_{\partial\Sigma}
d^{d-2}x\sqrt{\bg_{\partial\Sigma}} \,n_\mu
r_\nu\(\bxi_{\alpha}\bn^{\nu}\G_L^{\mu\alpha}-\bxi_{\alpha}\bn^{\mu}\G_L^{\nu\alpha}-\G_L^{\mu\alpha}\bn^{\nu}\bxi_{\alpha}+\G_L^{\nu\alpha}\bn^{\mu}\bxi_{\alpha}\) \,.
\eea
In asymptotically SdS spacetimes [see Eq.~(\ref{SdS})], the last two terms in Eq. (\ref{g charge}) fall off
too fast at large $r$ to contribute
 and the total energy is given by
 \be \label{energy last}
 E=\(N-\frac{4\,\Lambda(d-3)}{(d-1)(d-2)}N_1\) \frac{(d-2) \mathrm{Vol}(S^{d-2})}{2 }r_0^{d-3} \,,
 \ee
 or in the full explicit form as in Eq. (\ref{ENERGY FORMULA}).

\section{Discussion}
\label{conclusions}

In this paper, we have derived a simple formula \eqref{ENERGY
FORMULA} for the ADT energy of any gravitational theory of the form
$\sL = \sL(R^{\mu \nu}_{\rho \sigma}, g_{\mu \nu})$.  We gave a
detailed argument that the energy of such a theory takes the same
basic form as in quadratic curvature gravity, but with coefficients
modified by the higher-curvature terms.  The coefficients are given
by taking derivatives of the Lagrangian with respect to the Riemann
tensor, and in this sense our energy formula is reminiscent of
Wald's entropy formula.  We have demonstrated in a number of
examples that our formula correctly reproduces previous results, but
with significantly less computations.  For more complicated theories
in which following the full ADT procedure would be unmanageable in
practice, it seems that our formula could still be applied
relatively easily.

We note from the final formula for energy \eqref{ENERGY FORMULA}
that only $N$ and $N_{1}$ appear, and it would be interesting to
understand why this is the case. We also see that in $d=3$, the
contribution of the second derivative of the Lagrangian completely
drops out. In the case of three-dimensional topologically massive
gravity (TMG)\cite{Deser:1982vy,Deser:1981wh,Deser:1982sv}, the action
contains a gravitational Chern-Simons term so it is not of the form
$\sL = \sL(R^{\mu \nu}_{\rho \sigma}, g_{\mu \nu})$. Indeed, the ADT
energy for TMG has a different structure than \eqref{ENERGY FORMULA}
\cite{Deser:2003vh}.   It is also known that Wald's entropy formula
has to be modified in TMG, since the Chern-Simons term does not
satisfy the diffeomorphism-covariance requirement in the original
construction (see, e.g., Refs. \cite{Tachikawa:2006sz,Bonora:2011gz}).

Given the final expression for the energy (\ref{energy last}), it seems natural to define the effective gravitational coupling as
\be \frac{1}{2\kappa_{eff}}=N-\frac{4\,\Lambda(d-3)}{(d-1)(d-2)}N_1
\, .
\ee
Then the energy can be written succinctly in terms of the
Einstein gravity result as
\be E=\frac{\kappa}{\kappa_{eff}} E_{0}.
 \ee
This is analogous to the way the entropy was written
in Ref.~\cite{entropy} as \be S=\frac{A}{4\,G_{eff}} \, , \ee where $A$
is the black hole area. However, the effective coupling constant
also has an interpretation in the tree-level scattering amplitude
via the exchange of a graviton. When one looks at a similar process
on the background~\cite{Tekin}, the effective coupling turns out to
be the coefficient of $\G^{L}_{\mu\nu}$ in the stress-energy tensor:
\be \frac{1}{2\kappa_{eff}}=N-\frac{4 \Lambda}{d-2} N_1 -\frac{2
\Lambda}{(d-1)(d-2)} N_3 \,. \ee
The two definitions for the
effective coupling coincide for Lanczos-Lovelock gravity, since any
Lagrangian of the Lanczos-Lovelock type can be reduced to a Gauss-Bonnet
quadratic theory~\cite{Sisman:2012rc}. This coincidence might be
related to the fact that higher-derivative theories which are not of
the Lanczos-Lovelock type exhibit ghosts and other
inconsistencies~\cite{Brustein:2011,Brustein:2012}.  In future work,
it would be interesting to further understand this ambiguity in the
definition of the effective gravitational coupling.

\vspace{.5cm}\noindent
{\bf Note Added:}  One day after this paper appeared on the arXiv, the paper \cite{Senturk:2012yi} appeared with some overlapping material.

\section{Acknowledgments}

We thank Ramy Brustein for a discussion and Stanley Deser for comments.


\begin{thebibliography}{11}

\bibitem{wald1}
  R.~M.~Wald,
  ``Black hole entropy is the Noether charge,''
  Phys.\ Rev.\  D {\bf 48}, 3427 (1993)
  [arXiv:gr-qc/9307038].

\bibitem{wald2}
 V.~Iyer and R.~M.~Wald,
  ``Some properties of Noether charge and a proposal for dynamical black hole
  entropy,''
  Phys.\ Rev.\  D {\bf 50}, 846 (1994)
  [arXiv:gr-qc/9403028].

\bibitem{myers}
  T.~Jacobson, G.~Kang and R.~C.~Myers,
  ``On Black Hole Entropy,''
  Phys.\ Rev.\  D {\bf 49}, 6587 (1994)
  [arXiv:gr-qc/9312023].

\bibitem{Kastor}
D.~Kastor,
  ``Komar Integrals in Higher (and Lower) Derivative Gravity,''
  Class.\ Quant.\ Grav.\  {\bf 25}, 175007 (2008)
  [arXiv:0804.1832 [hep-th]].

\bibitem{Kastor2}
  D.~Kastor, S.~Ray and J.~Traschen,
  ``Smarr Formula and an Extended First Law for Lovelock Gravity,''
  Class.\ Quant.\ Grav.\  {\bf 27}, 235014 (2010)
  [arXiv:1005.5053 [hep-th]].

  \bibitem{shear1}
    R.~Brustein and A.~J.~M.~Medved,
    ``The Ratio of shear viscosity to entropy density in generalized theories of gravity,''
    Phys.\ Rev.\ D {\bf 79}, 021901 (2009)
    [arXiv:0808.3498 [hep-th]].

  \bibitem{shear2}
    M.~F.~Paulos,
    ``Transport coefficients, membrane couplings and universality at
   extremality,''
   JHEP {\bf 1002}, 067 (2010)
   [arXiv:0910.4602 [hep-th]].

  \bibitem{entropy}
    R.~Brustein, D.~Gorbonos and M.~Hadad,
    ``Wald's entropy is equal to a quarter of the horizon area in units of the effective gravitational coupling,''
    Phys.\ Rev.\ D {\bf 79}, 044025 (2009)
    [arXiv:0712.3206 [hep-th]].


  \bibitem{entropy2}
    R.~Brustein, D.~Gorbonos, M.~Hadad and A.~J.~M.~Medved,
    ``Evaluating the Wald Entropy from two-derivative terms in quadratic actions,''
    Phys.\ Rev.\ D {\bf 84}, 064011 (2011)
    [arXiv:1106.4394 [hep-th]].

  \bibitem{Tekin}
    T.~C.~Sisman, I.~Gullu and B.~Tekin,
    ``All unitary cubic curvature gravities in D dimensions,''
    Class.\ Quant.\ Grav.\  {\bf 28}, 195004 (2011)
    [arXiv:1103.2307 [hep-th]].

 \bibitem{Sisman:2012rc}
  T.~C.~Sisman, I.~Gullu and B.~Tekin,
  ``Spectra, vacua and the unitarity of Lovelock gravity in D-dimensional AdS spacetimes,''
Phys.\ Rev.\ D {\bf 86}, 044041 (2012)
  [arXiv:1204.3814 [hep-th]].


  \bibitem{Deser:2002rt}
    S.~Deser and B.~Tekin,
    ``Gravitational energy in quadratic curvature gravities,''
    Phys.\ Rev.\ Lett.\  {\bf 89}, 101101 (2002)
    [arXiv:hep-th/0205318].

  \bibitem{Deser:2002jk}
    S.~Deser and B.~Tekin,
    ``Energy in generic higher curvature gravity theories,''
    Phys.\ Rev.\  D {\bf 67}, 084009 (2003)
    [arXiv:hep-th/0212292].

  \bibitem{ADM}
    R.~L.~Arnowitt, S.~Deser, and C.~W.~Misner,
    ``Dynamical structure and definition of energy in general relativity,''
    Phys.\ Rev.\  {\bf 116}, 1322 (1959);

   ``Canonical variables for general relativity,''
    Phys.\ Rev.\  {\bf 117}, 1595 (1960);

    ``The dynamics of general relativity,'' in
  {\it Gravitation: an Introduction to Current Research}, L.~Witten
  ed. (Wiley 1962), pp 227-265.

  \bibitem{Abbott:1981ff}
    L.~F.~Abbott and S.~Deser,
    ``Stability of gravity with a cosmological constant,''
    Nucl.\ Phys.\  B {\bf 195}, 76 (1982).

  \bibitem{Balasubramanian:1999re}
    V.~Balasubramanian and P.~Kraus,
    ``A stress tensor for anti-de Sitter gravity,''
    Commun.\ Math.\ Phys.\  {\bf 208}, 413 (1999)
    [arXiv:hep-th/9902121].

  \bibitem{Henningson:1998gx}
    M.~Henningson and K.~Skenderis,
    ``The holographic Weyl anomaly,''
    JHEP {\bf 9807}, 023 (1998)
    [arXiv:hep-th/9806087].

  \bibitem{Hawking:1980gf}
    S.~W.~Hawking,
    ``The path integral approach to quantum gravity,''
    in {\it General Relativity:  An Einstein Centenary Survey}, eds. S.~W.~Hawking and W.~Israel, Cambridge University Press, Cambridge (1979).

  \bibitem{Barnich:2001jy}
    G.~Barnich and F.~Brandt,
    ``Covariant theory of asymptotic symmetries, conservation laws and  central
    charges,''
    Nucl.\ Phys.\  B {\bf 633}, 3 (2002)
    [arXiv:hep-th/0111246].

  \bibitem{Barnich:2007bf}
    G.~Barnich and G.~Comp\`ere,
    ``Surface charge algebra in gauge theories and thermodynamic integrability,''
    J.\ Math.\ Phys.\  {\bf 49}, 042901 (2008)
    [arXiv:0708.2378 [gr-qc]].

  \bibitem{Compere:2007az}
    G.~Comp\`ere,
    ``Symmetries and conservation laws in Lagrangian gauge theories with
    applications to the mechanics of black holes and to gravity in three
    dimensions,''
    arXiv:0708.3153 [hep-th].

  \bibitem{Lee:1990nz}
    J.~Lee and R.~M.~Wald,
    ``Local symmetries and constraints,''
    J.\ Math.\ Phys.\  {\bf 31}, 725 (1990).

  \bibitem{Wald:1999wa}
    R.~M.~Wald and A.~Zoupas,
    ``A General Definition of "Conserved Quantities" in General Relativity and
    Other Theories of Gravity,''
    Phys.\ Rev.\  D {\bf 61}, 084027 (2000)
    [arXiv:gr-qc/9911095].

  \bibitem{previous}
    A.~J.~Amsel and D.~Gorbonos,
    ``The Weak Gravity Conjecture and the Viscosity Bound with Six-Derivative Corrections,''
    JHEP {\bf 1011}, 033 (2010)
    [arXiv:1005.4718 [hep-th]].

 \bibitem{Hindawi:1995cu}
  A.~Hindawi, B.~A.~Ovrut and D.~Waldram,
  ``Nontrivial vacua in higher derivative gravitation,''
  Phys.\ Rev.\ D {\bf 53}, 5597 (1996)
  [hep-th/9509147].

 \bibitem{Gullu:2010em}
  I.~Gullu, T.~C.~Sisman and B.~Tekin,
  ``Unitarity analysis of general Born-Infeld gravity theories,''
  Phys.\ Rev.\ D {\bf 82}, 124023 (2010)
  [arXiv:1010.2411 [hep-th]].

  \bibitem{Gullu:2010vw}
  I.~Gullu, T.~C.~Sisman and B.~Tekin,
  ``All Bulk and Boundary Unitary Cubic Curvature Theories in Three Dimensions,''
  Phys.\ Rev.\ D {\bf 83}, 024033 (2011)
  [arXiv:1011.2419 [hep-th]].

\bibitem{Henneaux:1985tv}
    M.~Henneaux and C.~Teitelboim,
    ``Asymptotically Anti-De Sitter Spaces,''
    Commun.\ Math.\ Phys.\  {\bf 98}, 391 (1985).

\bibitem{LL}
L.~D.~Landau and E.~M.~Lifshitz, {\it The Classical Theory of
Fields}, Pergamon Press, Oxford (1975).

\bibitem{fierz-pauli}
  M.~Fierz and W.~Pauli
  ``On Relativistic Wave Equations for Particles
  of Arbitrary Spin in an Electromagnetic Field''
  Proc. R. Soc. Lond. A 1939 {\bf 173}, 211-232.

\bibitem{ortin}
T. Ortin, \emph{ Gravity and strings}, Cambridge University Press,
Cambridge 2004, p. 103-108.


  \bibitem{Deser:1982vy}
  S.~Deser, R.~Jackiw and S.~Templeton,
  ``Three-Dimensional Massive Gauge Theories,''
  Phys.\ Rev.\ Lett.\  {\bf 48}, 975 (1982).

  \bibitem{Deser:1981wh}
  S.~Deser, R.~Jackiw and S.~Templeton,
  ``Topologically Massive Gauge Theories,''
  Annals Phys.\  {\bf 140}, 372 (1982)
  [Erratum-ibid.\  {\bf 185}, 406 (1988)]
  [Annals Phys.\  {\bf 185}, 406 (1988)]
  [Annals Phys.\  {\bf 281}, 409 (2000)].

  \bibitem{Deser:1982sv}
  S.~Deser,
  ``Cosmological Topological Supergravity,''
  in {\it Quantum Theory Of Gravity, Essays in Honor of the 60th Birthday of Bryce C Dewitt}, edited by S. M. Christenson (Taylor and Francis, Jersey, 1984), p. 374.


\bibitem{Deser:2003vh}
  S.~Deser and B.~Tekin,
  ``Energy in topologically massive gravity,''
  Class.\ Quant.\ Grav.\  {\bf 20}, L259 (2003)
  [gr-qc/0307073].

  \bibitem{Tachikawa:2006sz}
  Y.~Tachikawa,
  ``Black hole entropy in the presence of Chern-Simons terms,''
  Class.\ Quant.\ Grav.\  {\bf 24}, 737 (2007)
  [hep-th/0611141].

  \bibitem{Bonora:2011gz}
  L.~Bonora, M.~Cvitan, P.~Dominis Prester, S.~Pallua and I.~Smolic,
  ``Gravitational Chern-Simons Lagrangians and black hole entropy,''
  JHEP {\bf 1107}, 085 (2011)
  [arXiv:1104.2523 [hep-th]].

\bibitem{Brustein:2011}
  R.~Brustein and A.~J.~M.~Medved,
  ``Non-perturbative unitarity constraints
  on the ratio of shear viscosity to entropy density
  in UV complete theories with a gravity dual,''
  Phys.\ Rev.\ D {\bf 84}, 126005 (2011)  [arXiv:1108.5347 [hep-th]].


  \bibitem{Brustein:2012}
  R.~Brustein and A.~J.~M.~Medved,
  ``Graviton n-point functions for UV-complete theories in Anti-de Sitter space,''
   Phys.\ Rev.\ D {\bf 85}, 084028 (2012)
   [arXiv:1202.2221 [hep-th]].


\bibitem{Senturk:2012yi}
  C.~Senturk, T.~C.~Sisman and B.~Tekin,
  ``Energy and Angular Momentum in Generic F(Riemann) Theories,''
Phys.\ Rev.\ D {\bf 86}, 124030 (2012)
  arXiv:1209.2056 [hep-th].

\end{thebibliography}
\end{document}